\documentclass[11pt,leqno,twoside]{article}

\usepackage{amsfonts,amsmath,amsthm,amssymb}
\usepackage{enumerate}
\usepackage{graphics,graphicx,subfigure}

\usepackage{color}
\usepackage{todonotes}
\usepackage{cancel}
\usepackage{url}
\usepackage{hyperref}
\usepackage{makeidx}
\usepackage{showidx}
\usepackage{multicol}        
\usepackage{xspace}
\usepackage{stmaryrd}        
\usepackage{pifont}          
\usepackage{fancybox}        
\usepackage{bm}

 \usepackage{fancyhdr}
\theoremstyle{plain}
 \theoremstyle{definition}
 \newtheorem{lem}{Lemma}
 \newtheorem{defn}[lem]{Definition}
 \newtheorem{thm}[lem]{Theorem}
 \newtheorem{prop}[lem]{Proposition}
 \newtheorem{cor}[lem]{Corollary}
 \newtheorem{notn}[lem]{Notations}
 \newtheorem{pb}[lem]{Problem}
 \newtheorem{form}[lem]{Formulae}
 
 \newtheorem*{rk}{Remark}
 \newtheorem*{com}{Comment}
 \newtheorem*{ex}{Example}
 \theoremstyle{remark}

 \newcommand{\blem}{\begin{lem}}
 \newcommand{\elem}{\end{lem}}
 \newcommand{\bdefn}{\begin{defn}}
 \newcommand{\edefn}{\end{defn}}
 \newcommand{\bthm}{\begin{thm} }
 \newcommand{\ethm}{\end{thm}}
 \newcommand{\bprop}{\begin{prop}}
 \newcommand{\eprop}{\end{prop}}
 \newcommand{\bcor}{\begin{cor}}
 \newcommand{\ecor}{\end{cor}}
 \newcommand{\bnotn}{\begin{notn}}
 \newcommand{\enotn}{\end{notn}}
 \newcommand{\bpb}{\begin{pb}}
 \newcommand{\epb}{\end{pb}}
 \newcommand{\bform}{\begin{form}}
 \newcommand{\eform}{\end{form}}
 \newcommand{\brk}{\begin{rk}}
 \newcommand{\erk}{\end{rk}}
 \newcommand{\bcom}{\begin{com}}
 \newcommand{\ecom}{\end{com}}
 \newcommand{\bex}{\begin{ex}}
 \newcommand{\eex}{\end{ex}}
 \newcommand{\bpf}{\begin{proof}}
 \newcommand{\epf}{\end{proof}}

\newcommand{\bR}{\mathbb{R}}

\newcommand{\be}{\begin{equation}}
\newcommand{\ee}{\end{equation}}
\newcommand{\bal}{\begin{align}}
\newcommand{\eal}{\end{align}}
\newcommand{\ba}{\begin{align*}}
\newcommand{\ea}{\end{align*}}
\newcommand{\bmx}{\begin{matrix}}
\newcommand{\emx}{\end{matrix}}
\newcommand{\bbmx}{\begin{bmatrix}}
\newcommand{\ebmx}{\end{bmatrix}}
\newcommand{\bpmx}{\begin{pmatrix}}
\newcommand{\epmx}{\end{pmatrix}}
\newcommand{\bvmx}{\begin{vmatrix}}
\newcommand{\evmx}{\end{vmatrix}}

\newcommand{\wt}{\widetilde}
\newcommand{\f}{\frac}
\newcommand{\df}{\dfrac}

\newcommand{\inc}{\subseteq}

\newcommand{\sgn}{\mathrm{sgn}}

\newcommand{\argmin}{{\rm argmin}\,}

\newcommand{\minimize}[1]{\underset{#1}{\rm minimize}\,}

\newcommand{\la}{\lambda}

\newcommand{\eps}{\varepsilon}
  
\newcommand{\TODO}[1]{{\color{red}\tiny TODO: {#1}}}

\title{\vspace{-20mm}The Sparsity of LASSO-type Minimizers\medskip\hrule height 1.2pt \vspace{-6mm}}
\author{Simon Foucart\footnote{S F. is partially supported by grants from the NSF (DMS-2053172) and from the ONR (N00014-20-1-2787).}  --- Texas A\&M University}
\date{\vspace{-6mm}\rule{100mm}{0.8pt}}

\newcommand\shorttitle{The Sparsity of LASSO-type Minimizers}
\newcommand\authors{S. Foucart}

\begin{document}
\maketitle

\vspace{-15mm}
\begin{abstract}
This note extends an attribute of the LASSO procedure to a whole class of related procedures, including
square-root LASSO,
square LASSO,
LAD-LASSO,
and an instance of generalized LASSO.
Namely,  under the assumption that the input matrix satisfies
an $\ell_p$-restricted isometry property (which in some sense is weaker than the standard $\ell_2$-restricted isometry property assumption),
it is shown that 
if the input vector comes from the exact measurement of a sparse vector,
then the minimizer of any such LASSO-type procedure has sparsity comparable to the sparsity of the measured vector.
The result
remains valid in the presence of moderate measurement error
when the regularization parameter is not too small.
\end{abstract}



\vspace{-5mm}
\begin{center}
\rule{100mm}{0.8pt}
\end{center}


\section{Introduction}

The purpose of this note is to extend a result established in \cite{FTZ},
namely that the LASSO output has sparsity at most proportional to the sparsity of the vector giving rise to the LASSO input.
More precisely, consider a vector $x \in \bR^N$ measured via $y =Ax \in \bR^m$
for some matrix $A \in \bR^{m \times N}$, $m \ll N$,
and suppose that $x$ is $s$-sparse, 
meaning that 
$$
\|x\|_0 := | {\rm supp}(x) | \le s,
\qquad \mbox{ where}   \quad
{\rm supp}(x) = \{ j \in [1 \colon N] : x_j \not= 0 \}.
$$
Let $x^\la \in \bR^N$ be the output of the LASSO procedure with regularization parameter $\la > 0$,
i.e.,  let $x^\la$ be a solution of
$$
\minimize{z \in \bR^N} \f{1}{2} \|y - Az\|_2^2 + \la \|z\|_1.
$$ 
Assume that the measurement matrix $A \in \bR^{m \times N}$ satisfies the standard restricted isometry property of order~$t$ (proportional to $s$) and constant $\delta \in (0,1)$,
meaning that 
\begin{equation}
\label{StRIP}
\sqrt{1-\delta} \, \|z\|_2 \le \|A z\|_2 \le \sqrt{1+\delta} \, \|z\|_2
\quad \mbox{whenever }\|z\|_0 \le t.
\end{equation}
Then, as shown in \cite{FTZ}, there is a constant $C_\delta$ depending on $\delta$ such that $\|x^\la\|_0 \le C_\delta \, s$.
For instance,  the choice $\delta=0.4$ guarantees that $\|x^\la\|_0 \le 4 s$.

In this note,
a similar result is established  for a generalization of the LASSO procedure branching out in four directions:
\vspace{-5mm}
\begin{itemize}
\item the power of the norm on $y-Az$ can be any $q \ge 1$---in particular, 
the case $q=1$ corresponds to the square-root LASSO (see \cite{BCW});
\item in fact, the norm on $y-Az$ is not restricted to be the $\ell_2$-norm, as any $\ell_p$-norm with $p \in [1,2]$  is possible---in particular, the case $p=1$ and $q=1$ corresponds to the least absolute deviation (LAD) LASSO (see \cite{HLJ});
\item the $\ell_1$-norm on $z$ can be raised to any power $r \ge 1$---in particular, the case $r=2$ corresponds to the square LASSO,
which can be recast as a nonnegative least squares problem (see \cite{FK});
\item the $\ell_1$-norm can act not only on $z$ but also on $B^{-1}z$
for some invertible matrix $B \in \bR^{N \times N}$---this corresponds to the generalized LASSO (see \cite{TT}, where $B^{-1}$ can instead be a rectangular matrix).
\end{itemize}

As in \cite{FTZ},
the full result incorporates measurement errors.
However, for simplicity,
a particularized version is stated here first.

\bthm
\label{ThmSimple}
Let $p \in [1,2]$, $q \ge 1$, and $r\ge 1$.
Consider a vector $x \in \bR^N$ such that $B^{-1} x$ is $s$-sparse
for some invertible matrix $B \in \bR^{N \times N}$ with condition number $\kappa_B := \|B\|_{2 \to 2}  \|B^{-1}\|_{2 \to 2}$.
Suppose that the vector $x$ is measured via $y = Ax \in \bR^m$ for some matrix $A \in \bR^{m \times N}$ satisfying the nonstantard restricted isometry property
of order $t = \lfloor (2 \gamma \kappa_B)^2  s \rfloor +1 $ with ratio $\gamma = \beta/\alpha$, i.e., 
$$
\alpha \|z\|_2 \le \|A z\|_p \le \beta \|z\|_2
\qquad \mbox{whenever }\|B^{-1}z\|_0 \le t.
$$
Then, for any $\la > 0$, the solution $x^\la$ of the LASSO-type procedure  
\begin{equation*}
\minimize{z \in \bR^N} \f{1}{q} \|y - Az\|_p^q  + \la \f{1}{r} \|B^{-1} z\|_1^r
\end{equation*}
has sparsity at most proportional to $s$, 
namely
$$
\|x^\la \|_0 \le \lfloor \chi^2 s \rfloor,
\qquad \chi :=  2\gamma \kappa_B.
$$
\ethm 

The restrictions $p,q,r \ge 1$ are here to make the optimization program convex.
The restriction $p \le 2$ is here to make the nonstandard restricted isometry property assumption realizable in the regime $m \asymp s \ln(eN/s)$.
This assumption is in fact weaker than the standard restricted isometry property assumption, as discussed in Subsection \ref{SubsecRIP}. 

The remainder of this note is dedicated  to the proof of Theorem~\ref{ThmSimple},
or rather of the more general version given in Theorem~\ref{ThmFull}.
The latter shows that the conclusion remains unchanged when the measurement vector $y =Ax + e$ features a nonzero error vector $e \in \bR^m$,
provided its magnitude $\|e\|_p$ is small enough compared to $\la$.
From a reversed point of view,  favored here,
it can also be claimed that the LASSO-type minimizer has sparsity at most proportional to $s$,
provided the regularization parameter $\la$ is large enough compared to $\|e\|_p$.

The organization is now straightforward:
Section \ref{SecPrel} establishes side results needed in the said proof
and Section \ref{SecMain} completes the arguments.

\section{Preliminary results}
\label{SecPrel}

This section collects some auxiliary results that will be invoked later.
The author is not aware of earlier appearances of these results,
but the vastness of the literature around compressive sensing makes such appearances not unlikely.

\subsection{Characterization of LASSO-type minimizers}

This subsection establishes a necessary and sufficient condition for a vector to be a solution of the general LASSO-type procedure.
This condition is essential in the subsequent considerations.
In the case of standard LASSO,
i.e., $p=2$, $q=2$, and $r=1$,
it seems to be folklore---it can be found e.g. in 
\cite[Section 3.4]{HTF},
\cite[Section 2.1]{TNV},
and \cite[Section 15.1]{Book}.
The arguments below could be shortcut by exploiting the notion of subdifferential,
but a more elementary proof has been preferred,  at the expense of brevity.

\bprop
\label{PropChara}
Let $p>1$, $q \ge 1$, and $r \ge 1$.
Given $y \in \bR^m$,  $A \in \bR^{m \times N}$,  and $B \in \bR^{N \times N}$,
the matrix $B$ being invertible,
a LASSO-type minimizer 
$$
x^\la = \underset{z \in \bR^N}{\argmin}  \f{1}{q} \|y - Az\|_p^q  + \la \f{1}{r}\|B^{-1} z\|_1^r
$$
is characterized, with $\nu^\la: =  \la \|y - A x^\la\|_p^{p-q} \|B^{-1}x^\la\|_1^{r-1}$ and $S^\la := {\rm supp}(B^{-1} x^\la)$, by
\begin{align}
\label{Cha1}
\phantom{\big|}  \big( B^\top A^\top \big( \sgn(y-Ax^\la) \, |y-Ax^\la|^{p-1} \big) \big)_j \phantom{\big|} 
& =  \nu^\la \sgn( (B^{-1}x^\la)_j),
& & j \in S^\la,\\
\label{Cha2}
\big| \big( B^\top A^\top \big( \sgn(y-Ax^\la) \, |y-Ax^\la|^{p-1} \big) \big)_\ell \big|
& \le  \nu^\la,
& & \ell \not\in S^\la.
\end{align}
\eprop

\bpf
Notice first  that
it is enough to consider the case $B = {\rm I}_N$ by setting $z' = B^{-1}z$ and $A' = AB$.
Second,  thanks to the convexity of the objective function,  temporarily denoted by $F$,  notice that a global minimizer $x^\la$ is equivalently a local minimizer,
so it is characterized by $ F(x^\la) \le F(x^\la + t u)$
for all $u \in \bR^N$ and all $t \in \bR$ small enough in absolute value.
Fixing $u \in \bR^N$,
one shall expand  $F(x^\la + t u)$ around $t=0$.
To start, with $I: = \{ i \in [1 \colon m]: (y - A x^\la)_i \not= 0 \}$,
one writes
\begin{align*}
\| y - A(x^\la +t u)\|_p^p
& = \sum_{i=1}^m | (y - A x^\la)_i - t (A u)_i |^p\\
& = \sum_{i \in I} | (y - A x^\la)_i  |^p \bigg( 1-t \f{(A u)_i}{(y-Ax^\la)_i} \bigg)^p
+ |t|^p  \sum_{i \not\in I} |  (A u)_i |^p.
\end{align*}
In view of $(1 - t \xi)^p = 1 - t p \xi + o(t)$
and of $|t|^p = o(t)$ since $p>1$, 
one obtains
\begin{align*}
\| y - A(x^\la +t u)\|_p^p
& =  \sum_{i \in I} | (y - A x^\la)_i  |^p \Big( 1-t p \f{(A u)_i}{(y-Ax^\la)_i}  \Big) + o(t)\\
& = \| y - A x^\la \|_p^p - t p \sum_{i \in I} \sgn((y - A x^\la)_i) |(y - A x^\la)_i|^{p-1} (Au)_i + o(t)\\
& = \| y - A x^\la \|_p^p - t p \sum_{i =1}^m \sgn((y - A x^\la)_i) |(y - A x^\la)_i|^{p-1} (Au)_i + o(t)\\
& = \| y - A x^\la \|_p^p - t p \langle A^\top \big( \sgn(y-Ax^\la) \, |y-Ax^\la|^{p-1} \big) , u \rangle + o(t),
\end{align*}
where the last-but-one step relied on $p>1$ through $|(y - A x^\la)_i|^{p-1}  = 0$ for $i \not\in I$.
It follows that 
\begin{align}
\nonumber
\| y - A(x^\la +t u)\|_p^q
& = \bigg[ \| y - A x^\la \|_p^p
\bigg( 1 -  t p \f{\langle A^\top \big( \sgn(y-Ax^\la) \, |y-Ax^\la|^{p-1} \big) , u \rangle}{\| y - A x^\la \|_p^p} + o(t)  \bigg)
 \bigg]^{q/p}\\
 \nonumber
& = \| y - A x^\la \|_p^q 
\bigg( 1 -  t q \f{\langle A^\top \big( \sgn(y-Ax^\la) \, |y-Ax^\la|^{p-1} \big) , u \rangle}{\| y - A x^\la \|_p^p} + o(t)  \bigg)\\
\label{Chara1}
& = \| y - A x^\la \|_p^q  - t q \|y-A x^\la\|_p^{q-p} \langle A^\top \big( \sgn(y-Ax^\la) \, |y-Ax^\la|^{p-1} \big) , u \rangle + o(t).
\end{align}
To continue,
 with $S^\la := \{ j \in [1 \colon N ]: x_j \not= 0\}$ (recall that $B = {\rm I}_N$ is assumed here),
one writes
\begin{align*}
\|x^\la + t u \|_1 & 
= \sum_{j \in S^\la} \sgn((x^\la + t u)_j) (x^\la + t u)_j 
+ |t| \sum_{\ell \not\in S^\la} |u_\ell|  = \sum_{j \in S^\la} \sgn(x^\la_j) (x^\la + t u)_j 
+ |t| \sum_{\ell \not\in S^\la} |u_\ell| \\
& = \|x^\la\|_1 + t \sum_{j \in S^\la} \sgn(x^\la_j) u_j + |t| \sum_{\ell \not\in S^\la} |u_\ell|,
\end{align*}
where the last-but-one step used the fact that, for $j \in S^\la$,
$\sgn((x^\la + t u)_j) = \sgn(x^\la_j)$ when $|t|$ is small enough.
It follows that
\begin{align}
\nonumber
\|x^\la + t u \|_1^r &
= \bigg[ \|x^\la\|_1  \bigg( 1 +  t \f{\sum_{j \in S^\la} \sgn(x^\la_j) u_j}{\|x^\la\|_1} + |t| \f{\sum_{\ell \not\in S^\la} |u_\ell|}{\|x^\la\|_1}   \bigg) \bigg]^r\\
\nonumber
& = \|x^\la\|_1^r  \bigg( 1 +  t r \f{\sum_{j \in S^\la} \sgn(x^\la_j) u_j}{\|x^\la\|_1} + |t| r \f{\sum_{\ell \not\in S^\la} |u_\ell|}{\|x^\la\|_1} +o(t)  \bigg)\\
\label{Chara2}
& = \|x^\la\|_1^r  + t r \|x^\la\|_1^{r-1} \sum_{j \in S^\la} \sgn(x^\la_j) u_j
+ |t| r \|x^\la\|_1^{r-1} \sum_{\ell \not\in S^\la} |u_\ell| + o(t).
\end{align}
According to \eqref{Chara1} and \eqref{Chara2},
the validity of $F(x^\la) \le F(x^\la + t u)$ for all $t$ small enough in absolute value is equivalent to
\begin{align*}
0 & \le t
\Big(- \|y-A x^\la\|_p^{q-p} \langle A^\top \big( \sgn(y-Ax^\la) \, |y-Ax^\la|^{p-1} \big) , u \rangle + \la \|x^\la\|_1^{r-1} \sum_{j \in S^\la} \sgn(x^\la_j) u_j
\Big)\\
& + |t| \Big( \la \|x^\la\|_1^{r-1} \sum_{\ell \not\in S^\la} |u_\ell|\Big) \\
& = t \Big\{ \sum_{j \in S^\la} \big( - \|y-A x^\la\|_p^{q-p}
\big( A^\top \big( \sgn(y-Ax^\la) \, |y-Ax^\la|^{p-1} \big) \big)_j  + \la \|x^\la\|_1^{r-1} \sgn(x^\la_j)
 \big) u_j  \Big\}\\
& + |t|  \Big\{ \sum_{\ell \not\in S^\la} \big( - \sgn(t) \sgn(u_\ell) \|y-A x^\la\|_p^{q-p}
\big( A^\top \big( \sgn(y-Ax^\la) \, |y-Ax^\la|^{p-1} \big) \big)_\ell + \la \|x^\la\|_1^{r-1} 
\big) |u_\ell| \Big\}
\end{align*}
for all $t$ small enough in absolute value.
The latter inequality holds for all such $t$'s and all $u \in \bR^N$
if and only if the first term in curly brackets equals zero for all $u$ supported on $S^\la$
and the second term in curly brackets is nonnegative for all $u$ supported on the complement of $S^\la$.
This first condition holds if and only if \eqref{Cha1} is satisfied 
and this second condition holds if and only if \eqref{Cha2} is satisfied.
The full characterization \eqref{Cha1}--\eqref{Cha2} of the LASSO-type minimizer has now been established in the sufficient case $B = {\rm I}_N$,
hence the proof is complete.
\epf

From the above characterization,
one easily sees that $x^\la = 0$ for
$\la \ge \|B^\top A^\top \big( \sgn(y) \, |y|^{p-1} \big)\|_\infty \|y\|_p^{q-p}$
in the usual case $r=1$.
In other words, the output of the LASSO-type procedure is as sparse as can be for $\la$ large enough.
In the case $r>1$, however,
$x^\la = 0$ would force $\nu^\la = 0$,
so \eqref{Cha1}-\eqref{Cha2} do not hold in general,
and the output of the LASSO-type procedure is not the zero vector (although it converges to it)
even for $\la$ large enough.
This remark reinforces the significance of the main result that the output of the LASSO-type procedure is sparse when $\la$ is large enough.

\subsection{Variation on the restricted isometry property}
\label{SubsecRIP} 

The standard restricted isometry property \eqref{StRIP} is not appropriate to the present setting,
where the sparsity does not concern the vector $x$ itself,
but $B^{-1}x$.
Thus, the restricted isometry property
\begin{equation}
\label{StRIP-D}
\sqrt{1-\delta} \, \|z\|_2 \le \|A z\|_2 \le \sqrt{1+\delta} \, \|z\|_2 
\qquad \mbox{whenever } \|B^{-1} z\|_0 \le t
\end{equation}
would be more appropriate.
This property---a special case of the $D$-restricted isometry property~\cite{CENR} where the dictionary $D$ is the invertible matrix $B$---holds for properly scaled subgaussian random matrices $A \in \bR^{m \times n}$ with failure probability at most $2\exp(-c \delta^2 m)$ provided
$m \ge C \delta^{-2} t \ln(eN/t)$.
Still, this restricted isometry property is not yet appropriate to the present setting,
which puts an $\ell_p$-norm on the measurement vectors $Az$.
Thus,
the nonstandard restricted isometry property utilized here reads 
\begin{equation}
\label{NonstRIPp}
\alpha \|z\|_2 \le \|A z\|_p \le \beta \|z\|_2
\qquad \mbox{whenever }\|B^{-1}z\|_0 \le t.
\end{equation}
It is to be noted that the constants $\alpha$ and $\beta$ need not be close to one---they could e.g.  scale with~$m$.
The only requirement is that the restricted isometry ratio $\gamma := \beta/\gamma$ is bounded by an absolute constant. 
For the main result to be nonvacuous,
it is important to point out that this nonstandard restricted isometry property holds for subgaussian matrices $A \in \bR^{m \times N}$
with failure probability at most $2\exp(-c m)$ provided
$m \ge C t \ln(eN/t)$, with constants $c,C$ depending on $\gamma$,  hence being absolute.
To see this, one could follow typical proofs of the $D$-restricted isometry property while making suitable modifications,
or realize that \eqref{NonstRIPp} is actually a consequence of the usual $D$-restricted isometry property,
as revealed by the observation below.
Thus, the nonstandard restricted isometry propery \eqref{NonstRIPp} is a weaker assumption than the standard restricted isometry property,
in the sense that it holds with overwhelming probability for at least as many ensembles as \eqref{StRIP-D}.
In fact,
it holds for more:
with $B={\rm I}_N$, the $\ell_1$-version of \eqref{NonstRIPp} holds for Laplace ensembles
in the optimal regime  $m \asymp t \ln(eN/t)$, see  \cite{FL},
while \eqref{StRIP-D} necessitates $m = \Omega( t \ln^2(eN/t))$,  see \cite{ALPT}.

\bprop
Given $p' \le p \le 2$,
if an ensemble of random matrices 
satisfies the $\ell_p$-version of~\eqref{NonstRIPp} with failure probability at most $2\exp(-c m)$ provided
$m \ge C t \ln(eN/t)$,
then it satisfies the $\ell_{p'}$-version of~\eqref{NonstRIPp} with failure probability at most $2\exp(-c' m)$ provided
$m \ge C' t \ln(eN/t)$.
\eprop

\bpf
Let $\theta \in (0,1)$ to be chosen soon. 
Assuming that $(1-\theta)m \ge C t \ln(eN/t)$,
the event that all the row-submatrices $A_I$, $|I| \ge (1-\theta)m$,
of a matrix $A \in \bR^{m \times N}$ from the random ensemble satisfy the $\ell_p$-version of~\eqref{NonstRIPp} occurs with failure probability at most
\begin{align*}
\sum_{k = (1-\theta)m}^m \binom{m}{k} \times 2 \exp(-ck)
& \le \sum_{\ell=0}^{\theta m} \binom{m}{\ell} \times 2 \exp(-c (1-\theta) m)
\le \bigg( \f{em}{\theta m} \bigg)^{\theta m} \times 2 \exp(-c (1-\theta) m)\\
& = 2 \exp\big( - ( c(1-\theta) - \theta \ln(e/\theta) ) m  \big).
\end{align*}
In this event, for any $z \in \bR^N$ with $\|B^{-1} z\|_0 \le t$,
one has $\|Az\|_{p'} \le m^{1/p'-1/p} \|Az\|_p$, so that
$$
\|Az\|_{p'} \le \beta' \|z\|_2,
\qquad \beta' := \beta m^{1/p' -1/p}. 
$$
Moreover,
according to Stechkin's estimate for the error of best sparse approximation in $\ell_p$ in term of the norm in $\ell_{p'}$,  see e..g.  \cite[Theorem 2.5]{Book},
there is some $I \inc [1 \colon m]$ of size $|I| = (1-\theta) m$ such that
$\|A_I z\|_p \le (1/(\theta m)^{1/p'-1/p}) \|A z\|_{p'}$,
Since $\|A_I z\|_p \ge \alpha \|z\|_2$, one derives that
$$
\|Az\|_{p'} \ge \alpha' \|z\|_2,
\qquad \alpha' := \alpha (\theta m)^{1/p' -1/p}. 
$$
Choosing the constant $\theta \in (0,1)$ such that $c(1-\theta) - \theta \ln(e/\theta) = c/2 =: c'$
and setting $C' = C/(1-\theta)$,
it has been proved that,  provided $m \ge C' t \ln(eN/t)$,
the $\ell_{p'}$-version of \eqref{NonstRIPp} holds with failure probability at most $2 \exp(-c' m)$,
while the restricted isometry ratio $\gamma ' = \beta'/\alpha'= \gamma (1/\theta)^{1/p'-1/p}$ is bounded by an absolute constant.
\epf

\subsection{Robust null space property from nonstandard restricted isometry property}

The robust null space property,
introduced in \cite[Section 4.3]{Book},
is a sufficient---and somewhat necessary---condition for robust sparse recovery by $\ell_1$-minimization (which corresponds to $\la=0$).
The standard robust null space property of order $s$ with constants $\rho \in (0,1)$ and $\tau > 0$, with respect to a norm $\|\cdot\|$ on $\bR^m$, reads:
$$
\|v_S\|_1 \le \rho \, \|v_{S^c}\|_1 + \tau \, \|Av\|
\qquad \mbox{for all } v \in \bR^N \mbox{ and } S \inc [1 \colon N]
\mbox{ with } |S|=s.
$$
It is well-known that the standard restricted isometry property \eqref{StRIP} implies this standard robust null space property (for $\|\cdot\| = \sqrt{s} \|\cdot\| _2$).
Here, it is shown that the nonstandard restricted isometry property \eqref{NonstRIPp} implies a nonstandard robust null space property order $s$ with constants $\rho \in (0,1)$ and $\tau > 0$,  which reads:
\begin{equation}
\label{RNSP}
\|(B^{-1}v)_S\|_1 \le \rho \, \|(B^{-1}v)_{S^c}\|_1 + \tau \, \sqrt{s} \|Av\|_p
\qquad \mbox{for all } v \in \bR^N \mbox{ and } S \inc [1 \colon N]
\mbox{ with } |S|=s.
\end{equation}
Special cases of the argument are routine,
see e.g.  \cite[Theorem 14.4]{BookDS} for the case  $p=1$ and $B = {\rm I}_N$.

\bprop
\label{PropRIPimpliesRNSP}
Given $p \ge 1$,
if a matrix $A \in \bR^{m \times N}$ satisfies the restricetd isometry property~\eqref{NonstRIPp} of order $t = \lceil ((1+\rho)\gamma \kappa_B / \rho)^2 s \rceil$ with ratio $\gamma = \beta/\alpha$,
then it satisfies the robust null space property~\eqref{RNSP} of order $s$ with constants $\rho$ and $\tau = (1+\rho) \|B^{-1}\|_{2 \to 2} \sqrt{s}/\alpha$ with respect to the $\ell_p$-norm on $\bR^m$.
\eprop

\bpf
Fixing $v \in \bR^N$,
note that it is enough to prove \eqref{RNSP} for an index set $S$ of $s$ largest absolute entries of $B^{-1}v$.
Let $T_0 \supseteq S$ be an index set of $t$ largest absolute entries of $B^{-1}v$,
and define also index sets $T_1,T_2,\ldots$ such that
\begin{align*}
T_1 & \mbox{ is an index set of next $t$ largest absolute entries of $B^{-1}v$},\\
T_2 & \mbox{ is an index set of next $t$ largest absolute entries of $B^{-1}v$},\\
\vdots
\end{align*}
For any $k \ge 1$,
the inequality $\|(B^{-1} v)_{T_k}\|_2 \le \|(B^{-1} v)_{T_{k-1}}\|_1/\sqrt{t}$
is easily obtained by comparing averages.
Now,  using (among other things) the leftmost inequality of \eqref{NonstRIPp},  
one can write
\begin{align*}
\|B (B^{-1} v)_{T_0} \|_2 & \le \f{1}{\alpha} \|A B (B^{-1} v)_{T_0} \|_p
= \f{1}{\alpha} \Big\| A B \Big (B^{-1}v - \sum_{k \ge 1} (B^{-1} v)_{T_k} \Big) \Big\|_p\\
& \le \f{1}{\alpha} \Big( \sum_{k \ge 1} \|A B (B^{-1} v)_{T_k} \|_p +  \|Av\|_p \Big).
\end{align*}
Next,  using (among other things) the rightmost inequality of \eqref{NonstRIPp}, 
one continues with
\begin{align*}
\|B (B^{-1} v)_{T_0} \|_2
& \le  \f{\beta}{\alpha}  \sum_{k \ge 1} \|B (B^{-1} v)_{T_k} \|_2 + \f{1}{\alpha} \|Av\|_p \\
& \le \gamma \|B\|_{2 \to 2}  \sum_{k \ge 1}  \|(B^{-1} v)_{T_k} \|_2  + \f{1}{\alpha} \|Av\|_p \\
& \le \gamma  \|B\|_{2 \to 2} \sum_{k \ge 1} \f{1}{\sqrt{t}} \|(B^{-1} v)_{T_{k-1}} \|_1  + \f{1}{\alpha} \|Av\|_p \\
& \le   \f{\gamma \|B\|_{2 \to 2}}{\sqrt{t}} \|B^{-1}v\|_1  + \f{1}{\alpha} \|Av\|_p  .
\end{align*}
Then, a bound on $\|(B^{-1} v)_{S} \|_1 $ is obtained via
\begin{align*}
\|(B^{-1} v)_{S} \|_1 
& \le \sqrt{s} \|(B^{-1} v)_{S} \|_2
\le \sqrt{s} \|(B^{-1} v)_{T_0} \|_2
\le \sqrt{s} \|B^{-1}\|_{2 \to 2} \|B (B^{-1} v)_{T_0} \|_2\\
& \le \gamma \kappa_B \sqrt{\f{s}{t}} \|B^{-1}v\|_1
+ \f{\sqrt{s} \|B^{-1}\|_{2 \to 2}}{\alpha} \|Av\|_p
 \le \f{\rho}{1+\rho} \|B^{-1}v\|_1
+ \f{\sqrt{s} \|B^{-1}\|_{2 \to 2}}{\alpha} \|Av\|_p,
\end{align*}
where the value $t = \lceil ((1+\rho)\gamma \kappa_B / \rho)^2 s \rceil$ was used in the last step. 
Finally, decomposing $\|B^{-1}v\|_1$ as $\|(B^{-1}v)_S\|_1 + \|(B^{-1}v)_{S^c}\|_1$ and rearranging, one deduces that
$$
\|(B^{-1} v)_{S} \|_1  \le \rho \|(B^{-1}v)_{S^c}\|_1 + \f{(1+\rho) \|B^{-1}\|_{2 \to 2} \sqrt{s}}{\alpha} \|Av\|_p,
$$
which is the announced robust null space property.
\epf

\section{The sparsity bound and its proof}
\label{SecMain}

In general,
without any assumption that $y \in \bR^m$ comes from a sparse vector,
one can guarantee that,  for (one of) the output(s) $x^\la$ of the LASSO-type procedure,  
the vector $B^{-1}x^\la$ is $m$-sparse.
Indeed, assuming uniqueness of $x^\la$ for simplicity,
if $S^\la = {\rm supp}(B^{-1}x^\la)$ had size $|S^\la| > m$,
then the columns of $AB \in \bR^{m \times N}$ indexed by $S^\la$ would be linearly dependent,
hence there would exist some $v \in \bR^N$ supported on $S^\la$ such that $ABv=0$.
Thus, for $u := Bv \in \bR^N$ and $t \in \bR$ small enough in absolute value,
the vector $x^\la + t u$ would be another LASSO-type minimizer,
in view of $A(x^\la + t u) = Ax^\la$ and of
\begin{align*}
\|B^{-1} (x^\la + t u)\|_1 & = \langle \sgn(B^{-1}(x^\la + t u)), B^{-1}(x^\la + t u) \rangle =  
 \langle \sgn(B^{-1}x^\la ), B^{-1}x^\la + t B^{-1} u \rangle\\
 & =  \|B^{-1} x^\la \|_1 + \f{t}{\nu^\la} \langle B^\top A^\top \big( \sgn(y-Ax^\la) \, |y-Ax^\la|^{p-1} \big) , B^{-1} u \rangle\\
& = \|B^{-1} x^\la \|_1 + \f{t}{\nu^\la} \langle \sgn(y-Ax^\la) \, |y-Ax^\la|^{p-1} , Au \rangle\\
& = \|B^{-1} x^\la \|_1.
\end{align*}

The main result below shows that,
under a favorable condition on $A \in \bR^{m \times N}$,
if $y \in \bR^m$ does come from a sparse vector,
then $B^{-1}$ applied to the output of the LASSO-type procedure in not simply $m$-sparse,
but $Cs$-sparse,
even with measurement error such that $\|e\|_p \le c \min\{ \|y\|_p, \la^{1/(q-r)} \}$.
The~constants $C$ and $c$, 
appearing under different names below,  have not been optimized.

\bthm
\label{ThmFull}
Let $p \in [1,2]$, $q \ge 1$, and $r\ge 1$.
Consider a vector $x \in \bR^N$ such that $B^{-1} x$ is $s$-sparse
for some invertible matrix $B \in \bR^{N \times N}$ with condition number $\kappa_B := \|B\|_{2 \to 2}  \|B^{-1}\|_{2 \to 2}$.
Suppose that the vector $x$ is measured via $y = Ax + e \in \bR^m$ 
for some error vector $e \in \bR^m$ with $\|e\|_p \le (1/3) \|y\|_p$
and some matrix $A \in \bR^{m \times N}$ satisfying the nonstantard restricted isometry property of order $t := \lfloor (6 \gamma \kappa_B)^2 s \rfloor +1 $ with ratio $\gamma = \beta/\alpha$, i.e., 
$$
\alpha \|z\|_2 \le \|A z\|_p \le \beta \|z\|_2
\qquad \mbox{whenever }\|B^{-1}z\|_0 \le t.
$$
Then, for any $\la \ge \la^* := 2^{q-1} \beta^r \|B\|_{2 \to 2}^r \|e\|_p^{q-r}$, 
the solution $x^\la$ of the LASSO-type procedure  
\begin{equation*}
\minimize{z \in \bR^N} \f{1}{q} \|y - Az\|_p^q  + \la \f{1}{r} \|B^{-1} z\|_1^r
\end{equation*}
has sparsity at most proportional to $s$, 
namely
$$
\|B^{-1} x^\la \|_0 \le \lfloor \chi^2 s \rfloor,
\qquad \chi :=  6 \gamma \kappa_B.
$$
\ethm

\bpf[Proof (Part I)]
The content of this part of the proof is valid for any $e \in \bR^m$,
but its conclusion only pertains to $e=0$.
The case $p=1$ is treated separately
because it is not covered by Proposition \ref{PropChara}. 

{\em The case $p>1$.}
The first step consists in bounding $\|B^{-1}x\|_1 - \|B^{-1}x^\la \|_1$ from above.
To do so,  according to Proposition~\ref{PropRIPimpliesRNSP} specified with $\rho=1$,
the robust null space property \eqref{RNSP} of order $s$ holds.
When applied to $v = x-x^\la$ and $S = {\rm supp}(B^{-1}x)$,
while taking $\| (B^{-1}v)_{S^c} \|_1 = \| (B^{-1}x^\la)_{S^c} \|_1$ and
$\| (B^{-1}v)_S \|_1 = \| B^{-1}x - (B^{-1} x^\la)_S \|_1 \ge  \| B^{-1}x\|_1 - \|(B^{-1} x^\la)_S \|_1$ into account,  it implies that $\| B^{-1}x\|_1 - \|(B^{-1} x^\la)_S \|_1 \le  \| (B^{-1}x^\la)_{S^c} \|_1
+ (2 \|B^{-1}\|_{2 \to 2} \sqrt{s}/\alpha) \|A(x-x^\la)\|_p$.
In other words, one has 
\begin{align}
\nonumber
\| B^{-1}x\|_1 - \|B^{-1} x^\la \|_1 & \le \f{2 \|B^{-1}\|_{2 \to 2} \sqrt{s}}{\alpha}  \|A(x-x^\la)\|_p\\
\label{Step1}
& \le \f{2 \|B^{-1}\|_{2 \to 2} \sqrt{s}}{\alpha} \big(  \|y - Ax^\la\|_p + \|e\|_p \big).
\end{align}
The second step consists in bounding $\|B^{-1}x\|_1 - \|B^{-1}x^\la \|_1$ from below.
To do so, one relies on \eqref{Cha1}--\eqref{Cha2}  to remark that $\| B^{-1} x^\la \|_1 = (1/\nu^\la) \langle B^\top A^\top \big( \sgn(y-Ax^\la) \, |y-Ax^\la|^{p-1} \big) , B^{-1}x^\la \rangle$
and that $\| B^{-1} x \|_1 \ge (1/\nu^\la) \langle B^\top A^\top \big( \sgn(y-Ax^\la) \, |y-Ax^\la|^{p-1} \big) , B^{-1}x \rangle$.
Thus, one deduces that
\begin{align}
\nonumber
\| B^{-1}x\|_1 - \|B^{-1} x^\la \|_1 
& \ge \f{1}{\nu^\la} \langle B^\top A^\top \big( \sgn(y-Ax^\la) \, |y-Ax^\la|^{p-1} \big) , B^{-1}x - B^{-1} x^\la \rangle\\
\nonumber
& =  \f{1}{\nu^\la} \langle  \sgn(y-Ax^\la) \, |y-Ax^\la|^{p-1}  ,  A(x - x^\la) \rangle\\
\nonumber
& = \f{1}{\nu^\la} \langle   \sgn(y-Ax^\la) \, |y-Ax^\la|^{p-1}  ,  y -A x^\la - e \rangle\\
\label{Step2}
& =  \f{1}{\nu^\la} \big(  \|y - A x^\la \|_p^p  -  \langle  \sgn(y-Ax^\la) \, |y-Ax^\la|^{p-1} ,   e \rangle \big).
\end{align}
By contradiction to the desired conclusion $\|B^{-1} x^\la\|_0 \le \lfloor \chi^2 s \rfloor$,
assume now that $S^\la := {\rm supp}(B^{-1} x^\la)$ has size $|S^\la| \ge \lfloor \chi^2 s \rfloor + 1$.
One can then consider a subset $T$ of $S^\la$ having size $|T| = \lfloor \chi^2 s \rfloor + 1$.
The vector $u := \big[B^\top A^\top \big( \sgn(y-Ax_\la) \, |y-Ax_\la|^{p-1} \big) \big]_T$ is then $t$-sparse and the nonstandard restricted isometry property yields
$\|ABu \|_p \le \beta \|Bu \|_2 \le \beta \|B\|_{2 \to 2} \|u\|_2$.
It follows that
\begin{align*}
\|u\|_2^2 & = \langle B^\top A^\top \big( \sgn(y-Ax_\la) \, |y-Ax_\la|^{p-1} \big), u \rangle 
=\langle  \sgn(y-Ax_\la) \, |y-Ax_\la|^{p-1} ,  A B u \rangle \\
& \le \|  \sgn(y-Ax_\la) \, |y-Ax_\la|^{p-1} \|_{p'} \|A B u  \|_p,
\end{align*}
where $p' = p/(p-1)$ is the conjugate exponent of $p$.
From the estimation of $\|ABu\|_p$ and the  observation that 
$\|  \sgn(y-Ax_\la) \, |y-Ax_\la|^{p-1} \|_{p'} = \| y -A x^\la \|_p^{p-1}$, 
the above inequality rearranges as
$  \| y -A x^\la \|_p^{p-1}  \ge (1 / (\beta \|B\|_{2 \to 2})) \|u\|_2$.
But according to \eqref{Cha1} and $T \inc S^\la$,
one has $\|u\|_2 = \nu^\la \sqrt{|T|}$,
and in turn
\begin{equation}
\label{Step3}
\| y -A x^\la \|_p^{p-1} 
\ge \f{\nu^\la \sqrt{|T|}}{\beta \|B\|_{2 \to 2}}.
\end{equation}

With all of this in place,  one can easily conclude in the case $e=0$---thus proving Theorem~\ref{ThmSimple}.
Indeed, \eqref{Step2} combined with \eqref{Step3} and with $|T| > \chi^2 s$ yields
$$
\| B^{-1}x\|_1 - \|B^{-1} x^\la \|_1 
> \f{\chi \sqrt{s}}{\beta \|B\|_{2 \to 2}}  \|y - A x^\la \|_p,
$$
while \eqref{Step1} reads
$$
\| B^{-1}x\|_1 - \|B^{-1} x^\la \|_1  
\le \f{2 \|B^{-1}\|_{2 \to 2} \sqrt{s}}{\alpha}  \|y - Ax^\la\|_p .
$$
Making the choice $\chi = 2 \gamma \kappa_B$,  one obtains a contradiction
to conclude that $\|B^{-1} x^\la\|_0 \le \lfloor \chi^2 s \rfloor$.

{\em The case $p=1$.}
A limiting argument does the trick here.
Fixing $q \ge 1$, $r \ge 1$, and $\la > 0$,
consider a sequence $(p_n)_{n \ge 1}$ of indices greater than $1$ but converging to $1$.
For each $n \ge 1$,
denote by $x^{(n)} \in \bR^N$ the LASSO-type minimizer corresponding to $p_n$,
which is now known to be $\lfloor \chi^2 s \rfloor$-sparse.
Note that the sequence $(x^{(n)})_{n \ge 1}$ is bounded,
as can be seen by evaluating the LASSO-type objective function at $z=0$ to write
$$
\la \f{1}{r} \| B^{-1}x^{(n)} \|_1^r
\le \f{1}{q}\|y - Ax^{(n)}\|_{p_n}^q + \la \f{1}{r} \| B^{-1}x^{(n)} \|_1^r
\le \f{1}{q}\|y \|_{p_n}^q \le \f{1}{q}\|y \|_{1}^q.
$$  
Therefore,  there is a subsequence $(x^{(n_k)})_{k \ge 1}$ extracted from $(x^{(n)})_{n \ge 1}$ and converging to some $x^* \in \bR^N$, say.
With $\wt{x} \in \bR^N$ denoting (one of) the LASSO-type minimizer(s) corresponding to $p=1$, one has, for any $k \ge 1$,
$$
 \f{1}{q}\|y - Ax^{(n_k)}\|_{p_{n_k}}^q + \la \f{1}{r} \| B^{-1}x^{(n_k)} \|_1^r
 \le \f{1}{q}\|y - A\wt{x}\|_{p_{n_k}}^q + \la \f{1}{r} \| B^{-1}\wt{x} \|_1^r,
$$
which passes to the limit as $k \to +\infty$ to give
$$
 \f{1}{q}\|y - Ax^*\|_{1}^q + \la \f{1}{r} \| B^{-1}x^* \|_1^r
 \le \f{1}{q}\|y - A\wt{x}\|_{1}^q + \la \f{1}{r} \| B^{-1}\wt{x} \|_1^r,
$$
showing that $x^*$ is a LASSO-type minimizer corresponding to $p=1$.
But, as a limit of $\lfloor \chi^2 s \rfloor$-sparse vectors, it must itself be a $\lfloor \chi^2 s \rfloor$-sparse vector.
\epf

Dealing with the  general situation where $e \not= 0$ is not much more complicated.
One shall use the following observation,  which may be folklore knowledge.

\blem
\label{LemInc}
The function $\la \mapsto \|y-Ax^\la\|_p$ is continuous (for $p>1$) and nondecreasing on $(0,+\infty)$,
with $\lim_{\la \to 0} \|y-Ax^\la\|_p \le \|e\|_p$
and $\lim_{\la \to +\infty} \|y-Ax^\la\|_p = \|y\|_p$.
\elem

\bpf
For $\mu > \la > 0$,
using the optimality of $x^\mu$ and the optimality of $x^\la$,
one can write
\begin{align*}
\f{1}{q} \|y - A x^\mu \|_p^q + \mu \f{1}{r} \|B^{-1} x^\mu \|_1^r
& \le \f{1}{q} \|y - A x^\la \|_p^q + \mu \f{1}{r} \|B^{-1} x^\la \|_1^r\\
& = \f{\mu}{\la} \bigg[ \f{1}{q} \|y - A x^\la \|_p^q + \la \f{1}{r} \|B^{-1} x^\la \|_1^r \bigg]
- \Big(  \f{\mu}{\la} -1 \Big) \f{1}{q} \|y - A x^\la \|_p^q \\
& \le \f{\mu}{\la} \bigg[ \f{1}{q} \|y - A x^\mu \|_p^q + \la \f{1}{r} \|B^{-1} x^\mu \|_1^r \bigg]
- \Big(  \f{\mu}{\la} -1 \Big) \f{1}{q} \|y - A x^\la \|_p^q.
\end{align*}
Rearranging the latter yields
$$
\Big(  \f{\mu}{\la} -1 \Big)\f{1}{q} \|y - A x^\la \|_p^q \le \Big(  \f{\mu}{\la} -1 \Big) \f{1}{q} \|y - A x^\mu \|_p^q,
$$
i.e.,  $ \|y - A x^\la \|_p \le  \|y - A x^\mu \|_p$ thanks to $\mu/\la-1 > 0$.
Thus, the monotonicity claim is proved.

Next, for $\la > 0$,
since the LASSO-type objective function at $x^\mu$ is no larger than at $x$ and at $0$,  one has
$$
\f{1}{q} \|y - A x^\la \|_p^q + \la \f{1}{r} \|B^{-1} x^\la \|_1^r
\le \left\{
\bmx
 \df{1}{q} \|e\|_p^q  +   \la \df{1}{r} \|B^{-1} x \|_1^r,\\
 \\
\df{1}{q} \|y\|_p^q.  \phantom{ + \la \df{1}{r} \|B^{-1} x \|_1^r,}
\emx
\right.
$$
The first inequality gives $\|y - A x^\la \|_p^q \le \|e\|_p^q +  (\la q/r) \|B^{-1} x \|_1^r$.
Letting $\la \to 0$ yields the claim about the first limit.
The second inequality gives $\|B^{-1} x^\la \|_1^r \le (r/(\la q)) \|y\|_p^q$.
Letting $\la \to + \infty$ yields $B^{-1} x^\la \to 0$, hence $x^\la \to 0$,
and the claim about the second limit follows. 

Finally, to establish the continuity for $p>1$,
consider a sequence $(\la_n)_{n \ge 1}$ of positive numbers converging to some $\la > 0$.
It is enough to show that the sequence $(A x^{\la_n})_{n \ge 1}$ converges to $A x^\la$.
By contradiction,
if this is not so,
then there exists $\eps>0$ and a subsequence $(A x^{\la_{n_k}})_{k \ge 1}$ such that $\|A x^{\la_{n_k}} - A x^\la \|_p \ge \eps$ for all $k \ge 1$.
By the boundedness of the sequence $(x_{n_k})_{k \ge 1}$ 
(recall that $\|B^{-1} x^{\la_{n_k}} \|_1^r \le (r/(\la_{n_k} q)) \|y\|_p^q$),
there is also a subsequence $(x^{\la_{n_{k_\ell}}})_{\ell \ge 1}$ converging to some $\wt{x} \in \bR^N$.
Passing to the limit as $\ell \to +\infty$ in $\|A x^{\la_{n_{k_\ell}}} - A x^\la \|_p \ge \eps$ shows in particular that $A \wt{x} \not= A x^\la$.
Passing to the limit in 
$$
\f{1}{q} \|y-Ax^{\la_{n_{k_\ell}}}\|_p^q + \la_{n_{k_\ell}} \f{1}{r} \|B^{-1}x^{\la_{n_{k_\ell}}} \|_1^r
\le
\f{1}{q} \|y-Ax^\la\|_p^q + \la_{n_{k_\ell}} \f{1}{r} \|B^{-1}x^\la \|_1^r
$$
also shows that $\wt{x}$ is a LASSO-type minimizer for the parameter $\la$,
and hence so is $(\wt{x} + x^\la)/2$.
But, by strict convexity of the $\ell_p$-norm for $p \ge 1$,
one deduces that $A \wt{x} = A x^\la$.
This provides the required contradiction to conclude the proof.   
\epf

\bpf[Proof of Theorem \ref{ThmFull} (Part II)]
To finish with the general situation where $e \not= 0$,
one separates the proof in three subparts:
the first two subparts assume $p>1$
and establish the existence of $\la^*$ before estimating it, 
and the third subpart treats the case $p=1$.

{\em Existence of $\la^*$.}
By the assumption $\|e\|_p \le (1/3) \|y \|_p$,
one notices that $2 \|e\|_p$ belongs to the interval $\big[ \lim_{\la \to 0} \|y-Ax^\la\|_p ,  \lim_{\la \to +\infty} \|y-Ax^\la\|_p \big]$,
so Lemma~\ref{LemInc} guarantees the existence of $\la^\flat >0$ such that $\|y - A x^{\la^\flat} \|_p = 2 \|e\|_p$.
Considering from now on any $\la \ge \la^{\flat}$,
Lemma~\ref{LemInc} also guarantees that
$$
\| y -A x^\la \|_p \ge 2 \|e\|_p,
\qquad \mbox{i.e.,} \quad
 \|e\|_p \le \f{1}{2} \| y -A x^\la \|_p.
$$
Similarly to Part I of the proof
but with $\chi$ to be chosen differently,
it is assumed by contradiction that $|S^\la| \ge \lfloor \chi^2 s \rfloor +1$
and one again picks some $T \inc S^\la$ with $|T| = \lfloor \chi^2 s \rfloor +1$.
One shall still exploit the inequalities \eqref{Step1}, \eqref{Step2}, and \eqref{Step3},
which remain valid.  
Starting from \eqref{Step2},  in view of
$$
\langle  \sgn(y-Ax^\la) \, |y-Ax^\la|^{p-1} ,   e \rangle
\le \| \sgn(y-Ax^\la) \, |y-Ax^\la|^{p-1}  \|_{p'} \|e\|_p
= \|y-A x^\la\|_p^{p-1} \|e\|_p,
$$
one obtains
\begin{align}
\nonumber
\| B^{-1}x\|_1 - \|B^{-1} x^\la \|_1 
& \ge \f{1}{\nu^\la} \|y - A x^\la \|_p^{p-1} \big(  \|y - A x^\la \|_p  - \| e\|_p  \big)\\
\nonumber
& \ge  \f{1}{\nu^\la} \|y - A x^\la \|_p^{p-1} \f{1}{2} \|y - A x^{\la}\|_p\\
\label{Step2b}
& \ge \f{\sqrt{|T|}}{2 \beta \|B\|_{2 \to 2}} \|y - A x^{\la}\|_p,
\end{align}
where \eqref{Step3} was used in the last step.
As for \eqref{Step1}, it now implies
\begin{equation}
\label{Step1b}
\| B^{-1}x\|_1 - \|B^{-1} x^\la \|_1 
\le \f{3 \|B^{-1}\|_{2 \to 2} \sqrt{s}}{\alpha}  \|y - Ax^\la\|_p.
\end{equation}
Because $|T| > \chi^2 s$,
the inequalities \eqref{Step2b} and \eqref{Step1b} become contradictory for the choice $\chi = 6 \gamma \kappa_B$. 
In summary,
for any $\la \ge \la^\flat$,
it has been proved that $\|B^{-1} x^\la\|_0 \le \lfloor \chi^2 s \rfloor$.
One can therefore select $\la^* = \la^\flat$ in the statement of Theorem~\ref{ThmFull}.

{\em Estimation of $\la^*$.}
The objective here is to show that $\la^\flat \le \la^\sharp$
for a more explicit $\la^\sharp$,
so that $\|B^{-1} x^\la\|_0 \le \lfloor \chi^2 s \rfloor$ holds whenever $\la$ is at least equal to $\la^\sharp$,
which can then be selected as $\la^*$.
Since it is now known that $s^\flat := \|B^{-1} x^{\la^\flat}\|_0 \le \lfloor \chi^2 s \rfloor$,
taking $T = {\rm supp}(B^{-1} x^{\la^\flat} )$ in \eqref{Step3} gives
$$ 
\|y-A x^{\la^\flat}\|_p^{p-1}
\ge \f{\nu^{\la^{\flat}} \sqrt{s^\flat}}{\beta \|B\|_{2 \to 2}}
= \f{\la^\flat \|y-Ax^{\la^\flat}\|_p^{p-q} \|B^{-1}x^{\la^\flat}\|_1^{r-1} \sqrt{s^\flat}}{\beta \|B\|_{2 \to 2}},
$$
which in turn yields 
\begin{equation}
\label{tempp}
\|y-A x^{\la^\flat}\|_p^{q-1}
\ge \f{\la^\flat  \|B^{-1}x^{\la^\flat}\|_1^{r-1} \sqrt{s^\la}}{\beta \|B\|_{2 \to 2}},
\qquad \mbox{hence} \quad
(2 \|e\|_p)^{q-1} \ge \f{\la^\flat  \|B^{-1}x^{\la^\flat}\|_1^{r-1}}{\beta \|B\|_{2 \to 2}}.
\end{equation}
From here,  one bounds from below the $\ell_1$-norm of $B^{-1} x^{\la^\flat}$ as
\begin{align*}
\nonumber
\|B^{-1} x^{\la^\flat} \|_1 & \ge \|B^{-1} x^{\la^\flat} \|_2
\ge \f{1}{\|B\|_{2 \to 2}} \|x^{\la^\flat}\|_2 \ge  \f{1}{\beta \|B\|_{2 \to 2}} \|A x^{\la^\flat}\|_p\\
& \ge \f{1}{\beta \|B\|_{2 \to 2}} (\|y\|_p - \| y -A x^{\la^\flat}\|_p)
\ge  \f{1}{\beta \|B\|_{2 \to 2}} ( 3\|e\|_p - 2 \| e \|_p)\\
& = \f{1}{\beta \|B\|_{2 \to 2}} \|e\|_p.
\end{align*}
Substituting the latter into \eqref{tempp},
one derives that $\la^\flat \le 2^{q-1} \beta^r \|B\|_{2 \to 2}^r \|e\|_p^{q-r}=:\la^\sharp$.
This justifies the announced expression for $\la^*$. 

{\em The case $p=1$.}
Since $\|e\|_1 \ge \|e\|_p$ for $p \in [1,2]$,
any regularization parameter $\la$ in the admissible range for $p=1$ is also in the admissible range for $p \in [1,2]$. 
The limiting argument of Part I then applies {\em mutatis mutandis},
since the $\lfloor \chi^2 s \rfloor$-sparsity of all the $x^{(n)}$'s is ensured.
\epf

\brk
The expression $\la^* = 2^{q-1} \beta^r \|B\|_{2 \to 2}^r \|e\|_p^{q-r}$ is certainly improvable, especially when~$q \le r$.
For instance, when $q=r$,  which includes square-root LASSO and square LASSO,
the threshold~$\la^*$ reduces to a positive quantity independent of the magnitude of $e$.
But this does not transition continuously to the case $e=0$,
where one can take $\la^*=0$, as revealed in Theorem~\ref{ThmSimple}.  
\erk

\end{document}